 \documentclass[pmlr,twocolumn,10pt]{jmlr} 

\usepackage[usestackEOL]{stackengine}
\usepackage{multirow}

\usepackage{booktabs}
\usepackage{siunitx}

\usepackage[switch]{lineno}

\usepackage{verbatim}

\theorembodyfont{\upshape}
\theoremheaderfont{\scshape}
\theorempostheader{:}
\theoremsep{\newline}

\jmlrvolume{259}
\jmlryear{2024}
\jmlrsubmitted{LEAVE UNSET}
\jmlrpublished{LEAVE UNSET}
\jmlrworkshop{Machine Learning for Health (ML4H) 2024} 

 \title[Barttender]{Barttender: An approachable \& interpretable way to compare medical imaging and non-imaging data}

\author{
 \Name{Ayush Singla} \Email{ayushsn@stanford.edu} \\
 \addr School of Computer Science, Stanford University \\
 \Name{Shakson Isaac} \Email{shakson\_isaac@g.harvard.edu}\\
 \Name{Chirag J Patel} \Email{chirag\_patel@hms.harvard.edu}\\
 \addr Department of Biomedical Informatics, Harvard Medical School
}

\begin{document}

\maketitle

\begin{abstract}

Imaging-based deep learning has transformed healthcare research, yet its clinical adoption remains limited due to challenges in comparing imaging models with traditional non-imaging and tabular data. To bridge this gap, we introduce Barttender, an interpretable framework that uses deep learning for the direct comparison of the utility of imaging versus non-imaging tabular data for tasks like disease prediction. 

Barttender converts non-imaging tabular features, such as scalar data from electronic health records, into grayscale bars, facilitating an interpretable and scalable deep learning based modeling of both data modalities. Our framework allows researchers to evaluate differences in utility through performance measures, as well as local (sample-level) and global (population-level) explanations. We introduce a novel measure to define global feature importances for image-based deep learning models, which we call gIoU. Experiments on the CheXpert and MIMIC datasets with chest X-rays and scalar data from electronic health records show that Barttender performs comparably to traditional methods and offers enhanced explainability using deep learning models.

\end{abstract}

\paragraph*{Data and Code Availability}
We used the CheXpert and MIMIC datasets, which contain chest X-rays and scalar data from electronic health records, for our analyses. These datasets are publicly available to other researchers through Stanford AIMI and PhysioNet. Our code is available at the following GitHub repository: \href{https://github.com/singlaayush/barttender}{github.com/singlaayush/barttender}.

\paragraph*{Institutional Review Board (IRB)}
This research is exempt from IRB approval as the analysis is based on secondary data which is publicly available.

\section{Introduction}
\label{sec:intro}

Imaging-based deep learning solutions have seen significant growth in healthcare research, with applications across various medical tasks \citep{alowais2023revolutionizing}. Despite this progress, many of these solutions have not transitioned into clinical practice \citep{kelly2019key}. One key challenge in achieving this transition lies in the difficulty of comparing state-of-the-art imaging tools with the current clinical standards, which are often measured using non-imaging and tabular-based modalities.

This paper introduces Barttender, an approach designed to compare medical imaging data with non-imaging, tabular data in an approachable and interpretable manner. The Barttender methodology transforms non-imaging features—such as age, race, and Body Mass Index (BMI)—into grayscale bars that are appended to the corresponding medical images. This transformation allows for a direct comparison of the performance of deep learning (DL) models across imaging and tabular data. This is in contrast to the typical approach of converting images to scalars, which limits comparison to shallow machine learning models like logistic regression and XGBoost. Barttender also offers a robust explainability framework that aggregates attribution maps to assess the relative importance of imaging and non-imaging features at scale.

We demonstrate the potential of Barttender through a proof-of-concept evaluation using two widely-used medical imaging datasets: CheXpert and MIMIC. We have preliminary data to show that our approach allows simultaneous visualization of both data types and, for the case studies, has predictive capability equivalent to common tabular approaches, such as logistic regression.  

Our approach provides a way to not only compare prediction measures, but glean feature importances at both the sample and global level. Typically, image attribution approaches like saliency maps only output information on the sample level, for a single image. Barttender allows us to aggregate these attributions to derive a global view of feature importances over the entire dataset. These feature importances are conceptually analogous to a “coefficient” in a regression model. Moreover, our approach is robust to missing data, a critical factor in medical settings where data missingness is common. A review of related works, highlighting similar approaches and detailing how our method advances beyond them, is provided in Appendix \ref{apd:related_works}.

Our experiments on the CheXpert and MIMIC datasets demonstrate that Barttender performs comparably to the most commonly used frameworks for comparing imaging and non-imaging data, where shallow models like XGBoost are trained on scalars derived from images. The explainability module reveals that the model learned robust features from the tabular data encoded as bars. Additionally, Barttender recaptitulated what is known about some diseases,  such as highlighting the important role that certain non-imaging features played in predicting conditions like Cardiomegaly even when imaging data was available. 

Barttender helps researchers explain whether image modalities drive prediction independently from tabular risk factors (i.e., non-image data). Separating the influence of risk factors from image features is a challenge in current DL multi-modal paradigms. To address this, we introduce a gIoU (global 'IoU') measure in Barttender for assessing global feature importances. Moreover, our sample-level ‘IoU’ allows researchers to focus on whether specific case examples are influenced by tabular risk factors. Thus, the Barttender framework can facilitate the presentation of medical imaging-based deep learning approaches to clinical stakeholders.
 
\begin{figure*}[hbtp]
\floatconts
  {fig:methodology}
  {\caption{Barttender overview. First, non-imaging tabular EHR features are converted into grayscale bars. Second, these bars are appended to medical images and blank images to create Image and Blank Barttenders respectively. Third, image-based deep learning models are trained on these Barttenders. These models can be used to evaluate differences in utility between EHR and imaging data for a task through performance measures (e.g., ROC curves), as well as, both local, sample-level explanations (e.g., attribution maps) and global, population-level explanations (via gIoU).}}
  {\includegraphics[width=0.8\textwidth]{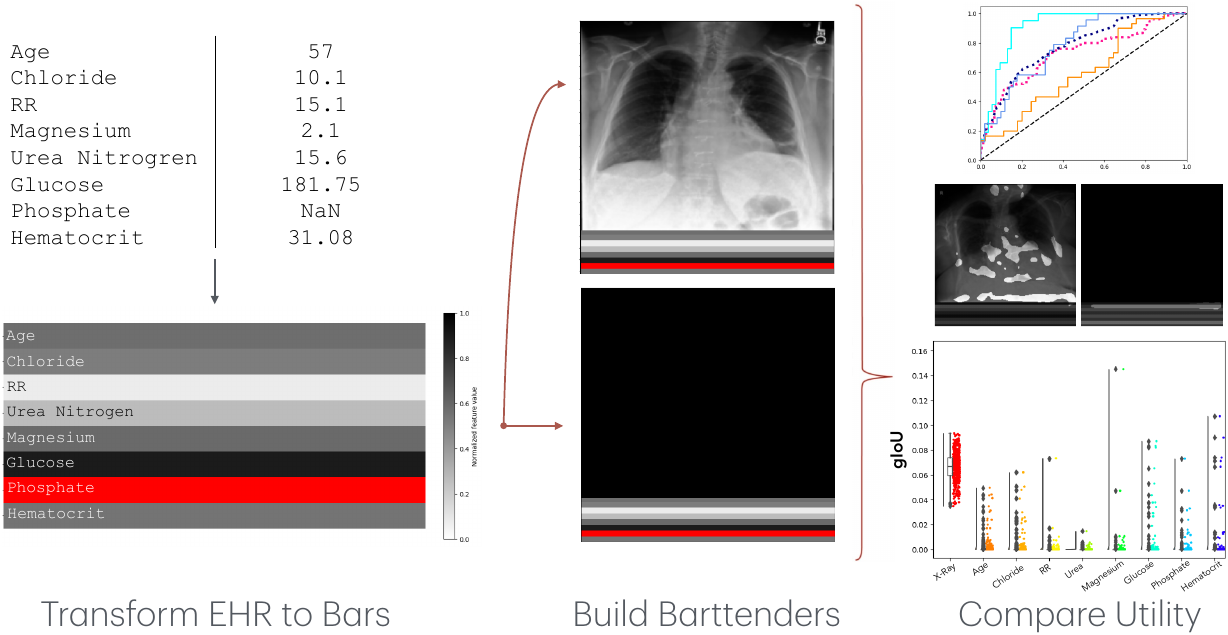}}
\end{figure*}

\section{Methodology}
\label{sec:methods}

The \textit{Barttender} methodology involves transforming tabular data into a set of grayscale bars, which are then appended to corresponding medical images to create a dataset of combined images called the \textit{Image Barttender}. A ``control" dataset, known as \textit{Blank Barttender}, is created by appending the same set of bars to blank images with dimensions matching the medical image.

Separate convolutional neural network models are trained on the \textit{Image Barttender} and \textit{Blank Barttender}. Using these models, relative performance is compared to evaluate medical images against non-imaging features.

Assessing the relative contribution of imaging and non-imaging data for prediction involves 3 steps. First, attribution maps are extracted for each test image. Next, a mean attribution intensity scaled by Intersection over Union (IoU) is computed to quantify local feature importance for each image. Last, the mean attribution intensity scaled by IoU is averaged across the test images to calculate a global feature importance metric called \textit{gIoU}. 

\subsection{Transforming non-imaging data into grayscale bars}

Each non-imaging variable are transformed into a grayscale bar, as shown in Figure \ref{fig:methodology}. To achieve this, values for each variable are normalized to a $[0, 1]$ range. Each normalized value is mapped to a grayscale gradient, where 0 corresponds to white and 1 to black. To easily discriminate values through visual inspection, custom rules are applied to each feature to normalize values between $0$ to $1$.

\paragraph{Continuous variables}
Continuous variables are quantile normalized. This technique minimizes the impact of outliers and distributional differences by aligning values based on their rank, rather than absolute values. This transformation makes variables with different distributions comparable. Since age (measured in years) converts to a $[0, 1]$ range when divided by $100$, we simply divide age by $100$. If any patients have age greater than 100, we set the value to 1. 

\paragraph{Categorical variables}
For binary categorical variables (e.g., yes/no) like sex defined at birth, values are set far apart on the gradient (e.g., 0.2 for No and 1.0 for Yes) to allow for easy visual discrimination.

For multi-categorical variables like race/ethnicity, values are assigned equidistant from one another in the $[0, 1]$ range (e.g., for three race options, assign 0.33, 0.66, and 0.99, respectively) to ensure maximum visual discrimination. 

However, if some categories are conceptually closer together than others, a custom scale may be appropriate. For instance, with variables like smoking status with options ``never smoked," ``former smoker," and ``current smoker". The values ``never smoked" and ``former smoker" could be assigned values closer together (e.g., 0.1 and 0.4), while ``current smoker" is set further apart (e.g., 0.9) to reflect a more significant difference.

\paragraph{Missing values}
 We propose representing missing values across variables using a red bar. This uniform representation for missing values across variables will have no overlap with other bars (in grayscale), allowing the deep learing model to detect patterns in missing data and evaluate potential bias in predictions for samples with missing data.

\subsection{Image and Blank Barttenders}

The processed bars are concatenated with medical images to create the Image Barttender, as shown in Figure \ref{fig:methodology}. This concatenation can be performed either horizontally or vertically, with the preferred orientation being the one that best preserves the original aspect ratio of the medical image. 

To evaluate the utility of the scalar variables which were converted to bars themselves, we also conduct a "Blank Barttender" ablation, as shown in Figure \ref{fig:methodology}. In the Blank Barttender ablation, the processed bars are concatenated with a blank image (of the same size as the medical image) similar to the Image Barttender. The blank image represents the exclusion of image data. This dataset of images, with information from only non-imaging data, is trained on the same deep learning model as Image Barttender. 

Thus, the model's performance on Blank Barttenders is comparable to other models trained solely on non-imaging data. Since the processed bars and model parameters remain unchanged, we can directly compare the performance of Image and Blank Barttenders. This comparison allows us to understand the added utility of using images for the task at hand (for e.g., disease prediciton). 

\subsection{Explainability using image attributions}

To enhance our understanding of the features on which the model focuses on each Barttender, we compute a global feature importance metric, gIoU, across the test set by aggregating attribution maps. Attribution maps, also called saliency maps, are a popular approach to creating post-hoc explanations for deep learning predictions on images \citep{simonyan2013deep}. These methods produce estimates of the relevance of each pixel of an image to the model output, which can be displayed as a mask over the input image that highlights important pixels. Since approaches to extract attribution maps are specific to deep learning architectures, they can be selected as appropriate for the experiment design with our methodology. However, these approaches can be unreliable and inconsistent at times \citep{tomsett_sanity_2020}. Thus, we recommend extracting attribution maps using more than one approach to validate the results observed. In our experiments, we used two such approaches — gradient-scaled saliency maps \citep{simonyan2013deep} and Integrated Gradients \citep{sundarajan_axiomatic_2017}.

\paragraph{Quantifying feature importance using gIoU}

Attribution maps produce estimates of the relevance of each pixel to an image to the model output (also described as producing estimates the attention paid by the model to each pixel). We recommend applying a minor Gaussian blur to these estimates to average pixel relevance scores (pixel intensities) across neighboring pixels. Since a pixel derives meaning in the context of its neighborhood, applying a Gaussian blur reduces any sharp distinctions between adjacent pixels and allows for significant, larger-scale areas to be highlighted in the attribution map. Following this blur, a small fraction (say, 10\%) of the pixels with the highest intensities in the attribution map are selected and overlaid as a mask on top of the image. 

To aggregate these intensities across the different regions in the Barttender image, one could simply report the mean pixel intensity across the most important pixels of the image that occur in a specific region; however, mean pixel intensity is not sufficient to help quantify feature importance for a single image. Consider the case where almost all of the model attention is focused on the medical image, and the remaining attention is focused on one of the Barttender bars. If we simply considered the mean pixel intensity among the most important pixels in both regions as the feature importance measure, the feature importance would be similar for both regions if their mean intensities were similar. This evaluation would be incorrect, since the medical image receives the majority of the area of the image under attention from the model. Therefore, the area of the attention on the Barttender region must be taken into consideration here. 

In addition to the area of model attention on the Barttender region, the area of the Barttender region itself must also be considered. If 10\% of the model attention is focused on some Barttender region, the area of the region helps us understand the relative importance of this attention. If this attention is focused on a bar, then it could be significant, since this attention might account for a large fraction of the bar’s region. However, if the region in question is the medical image, the attention might not be as significant, since the model focuses on only a small fraction of the image.

We take both the area of model attention and the area of the Barttender region into account by using their Intersection over Union (IoU) to scale the mean pixel intensity in the Barttender region. We define the IoU similarly to that in the image detection and segmentation literature, by computing the ratio of the intersection and the union of the two areas, i.e., A and B, in question:

\[ \texttt{IoU} = \frac{A \cap B}{A \cup B} \]

If we set the area of the model attention across the entire Barttender image as $A$, and the area of the Barttender region as $B$, $A \cap B$ represents the area of model attention across the Barttender region, and $A \cup B$ represents the union of the area of the Barttender region and the area of the model attention outside  the Barttender region. Thus, the IoU allows us to account for the area of model attention inside the Barttender region and the area of the entire Barttender region into a single scalar, which can be used to weight the mean pixel intensity, $\bar{p}$. 

\[ \texttt{gIoU} = \frac{1}{|T|} \sum \bar{p} * IoU \]

Averaged across the amount of samples in the test set, T, the gIoU is the global feature importance metric.

We believe gIoU can help researchers understand the role of different covariates in model predictions. For example, if the gIoU of the image in a Barttender is high, it would indicate that the image drives the prediction rather than the tabular covariates.

\section{Experiments}
\label{sec:experiments}

\subsection{CheXpert}

Our CheXpert experiments provide a proof-of-concept of our methodology. Our CheXpert sample \citep{irvin_chexpert_2019} derived from \citep{glocker_algorithmic_2023} contains a total of 42,884 patients with 127,118 chest X-ray scans with labels for 14 different conditions. The sample came pre-split into three sets for training (76205; 60\%), validation (12673; 10\%), and testing (38240; 30\%). Detailed patient demographics — specifically, age, self-reported racial identity, and biological sex — were available for these chest X-ray scans \citep{banerjee_reading_2022}. However, no other clinical information was available. Thus, in our experiments with CheXpert, we selected the demographic variables, i.e., age, race, and sex, as the tabular features to compare with the medical images, i.e., the chest X-rays. 

We expected that images would be preferred over demographics by any deep learning model for predicting chest conditions using the CheXpert dataset. Previous works have obtained state-of-the-art performance on CheXpert for the multi-label classification for the 14 different conditions described in the dataset using deep learning approaches applied only to Chest X-rays \citep{irvin_chexpert_2019}. Moreover, demographic variables alone contain little predictive information for the conditions annotated in CheXpert. Through our CheXpert experiments, we assessed whether this expectation is replicated by Barttender, and evaluated the capability of our Barttender methodology to successfully represent information from tabular features via comparisons with logistic regression and order ablations. 

Our CheXpert experiments had three parts.

First, we compared the deep learning model’s performance on the Blank Barttender to a logistic regression model trained on the tabular variables. If their performance was comparable, it would support our hypothesis that our bar-based representation can capture signal equivalent to the logistic regression.

Second, we compared the performance of the Image Barttender and the Blank Barttender, where we expected to see improved performance when the X-ray was included, as a sanity check.

Third, we assessed whether the model’s attribution to the bars is genuine and not spurious for both X-ray and Blank Barttenders. To test this, we analyzed how model performance changed as the order of the bars was varied. We conducted this analysis across all six possible orderings of age, race, and sex after transforming them to the $[0,1]$ range to create Barttenders as outlined in Section \ref{sec:methods}.

\subsection{MIMIC}

To evaluate the effectiveness of our methodology in this proof-of-concept, we compared our approach with the most commonly-used framework for comparing images and tabular data in medicine. As mentioned in Section \ref{sec:related_works}, this framework compares imaging and non-imaging data by developing image-derived scalar biomarkers. We compared our Barttender approach with this oft-used framework by utilizing the image-derived cardiomegaly biomarker values developed by \cite{duvieusart_multimodal_2022} for the MIMIC dataset \citep{johnson_mimic-iv_nodate, johnson_mimic-cxr-jpg_nodate}. 

\cite{duvieusart_multimodal_2022} extracted two image biomarkers using image detection and segmentation algorithms, namely the cardiothoracic ratio (CTR) and cardiopulmonary area ratio (CTAR), specifically for cardiomegaly — a condition where the heart is abnormally enlarged. The authors also evaluated the performance gain offered by these biomarkers when integrated with non-imaging, tabular data for cardiomegaly prediction. They evaluated XGBoost models trained on tabular clinical data from a small subset of the MIMIC dataset ($\approx 2700$ samples), both including and excluding these image biomarkers. The performance of the XGBoost model improved when the image biomarkers were included, indicating that images contain predictive information for cardiomegaly that is not captured by the tabular data. Interestingly, the tabular features alone demonstrated strong performance, indicating that they hold valuable information for predicting cardiomegaly, even without the inclusion of image biomarkers.

This setting in the MIMIC dataset provided us with an excellent opportunity to assess the efficacy of our approach. If effective, we expected to see comparable performance between the Blank Barttender and the XGBoost model trained solely on the tabular features. In addition, we expected to observe an improvement in performance with the Image Barttender compared to the Blank Barttender, reflecting the enhancement observed when image biomarkers were included in \cite{duvieusart_multimodal_2022}. 

Our MIMIC sample consists of $2667$ multi-modal (imaging + tabular) samples containing a chest X-ray image, two cardiomegaly image biomarkers, and tabular features. These tabular features include $14$ laboratory tests, $14$ ICU chart/vital sign values, and patient metadata such as demographics. This sample combines data from the MIMIC-IV and MIMIC-CXR-JPG datasets \citep{johnson_mimic-iv_nodate, johnson_mimic-cxr-jpg_nodate}, preprocessed as directed by \cite{duvieusart_multimodal_2022} (details can be found in the appendix, section \ref{sec:mimic_prep}).

\paragraph{Feature Selection} We extracted $ \approx $ 22,000 samples for which the non-imaging variables were available in the MIMIC dataset to select the subset of variables we would use to create bars. We used this larger sample for feature selection to minimize bias. We excluded the variables for which over half the samples had missing data from this analysis. This exclusion was necessary to ensure comparability of our approach to logistic regression, which cannot accommodate for missing data. We ran logistic regression after performing mean imputation on the remaining variables to identify the feature importance among these variables for predicting Cardiomegaly. 8 tabular features were found to have statistically significant Z-scores, and were selected for use in all our MIMIC experiments.

The 8 selected non-imaging variables, which consist of age, vital signs, and laboratory values, are all continuous. Age was processed to the $[0,1]$ range and quantile normalization was applied to the remaining variables, as described in Section \ref{sec:methods}. These variables were used to create the blank and Image Barttender, and to train XGBoost and logistic regression models for a fair comparison. 

We used the XGBoost model from \cite{duvieusart_multimodal_2022}. To ensure a fair comparison with deep learning models, we optimized the learning rate, maximum tree depth, and tree sub-sample through a grid search for the XGBoost model. We used early stopping to ensure that we optimally tuned the number of decision trees.

We used indicator variables for the logistic regression model, and the indicator variable had an insignificant z-score for all experiments. Mean imputation was applied to handle missing data before training for only the logistic regression models. Both XGBoost and Barttenders are capable of modeling missing data.

\section{Results}
\label{sec:results}

Our main results showcasing the strength of our approach can be found in Tables \ref{tab:chexpert-metrics} and \ref{tab:mimic-metrics}, and are summarized below. The deep learning models trained on our Blank and Image Barttenders are comparable to XGBoost and logistic regression models trained on scalar representations of imaging and non-imaging features in both MIMIC and CheXpert. The CheXpert experiments show that model attention for Barttenders is order-invariant, while the $gIoU$ values computed for MIMIC reveal interesting insights into training dynamics. These findings were largely reflected by both gradient-based saliency maps and Integrated Gradients, indicating that our findings here could be generalizable.

\begin{table}[htbp]
\setlength{\tabcolsep}{3pt}
\floatconts
  {tab:chexpert-metrics}
  {\caption{Performance metrics of different approaches trained on patient demographics for selected labels of CheXpert. All reported metrics are averaged across 10-folds; Barttender metrics are additionally averaged over all six order permutations of age, race and sex. Metrics for all $14$ labels with CIs are reported in Table \ref{tab:chexpert-metrics-all}. As expected, Image Barttenders significantly outperform Blank Barttenders.}}
{\begin{tabular}{llccc}
\toprule
\bfseries Method    & \bfseries Label               & \bfseries AUC & \bfseries F1 & \bfseries MCC  \\
\midrule
    Blank Bartt.    & Cardiomegaly                  & 0.53  & 0.48 & 0.10  \\
    Blank Bartt.    & Pleural Effusion              & 0.53  & 0.49 & 0.06  \\
    Blank Bartt.    & Pneumonia                     & 0.50  & 0.43 & 0.00  \\
    Log. Reg.       & Cardiomegaly                  & 0.57  & 0.38 & 0.07  \\
    Log. Reg.       & Pleural Effusion              & 0.53  & 0.34 & 0.05  \\
    Log. Reg.       & Pneumonia                     & 0.50  & 0.20 & -0.01 \\
    X-ray Bartt.    &   Cardiomegaly                & 0.80 & 0.80 & 0.61 \\
    X-ray Bartt.    &   Pleural Effusion            & 0.82 & 0.81 & 0.64 \\
    X-ray Bartt.    &   Pneumonia                   & 0.67 & 0.68 & 0.42 \\

\bottomrule
\end{tabular}}
\end{table}

\subsection{CheXpert}

\paragraph{Bars reflect signal in tabular data} Table \ref{tab:chexpert-metrics} shows that the DenseNet-121 model trained on the Blank Barttenders has predictive performance similar to that of a logistic regression model trained on these features. Consider the Matthews correlation coefficient (MCC) here, a measure that considers all four categories of a confusion matrix whose value ranges from -1 to 1, where 1 indicates a perfect prediction, 0 indicates no better than random guessing, and -1 indicates completely incorrect prediction \citep{chicco2023matthews}. Logistic regression performs minimally better than random for Cardiomegaly and Pleural Effusion, and negligibly worse than random for Pneumonia. 

\begin{figure}[htbp]
\floatconts
  {fig:xray}
  {\caption{Raincloud plot, which combines a scatter, box, and density plot, of gIoUs for Image Barttenders computed using gradient-based saliency maps.}}
  {
      \includegraphics[width=\linewidth]{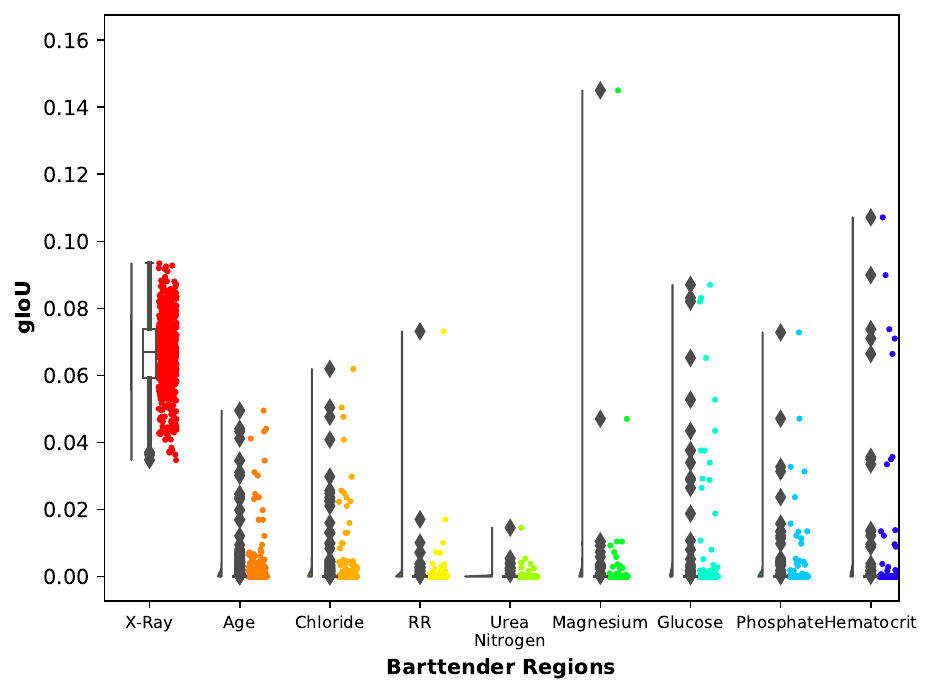}
  }
\end{figure}

This relative ranking of performance across these conditions is replicated by the Blank Barttender, indicating that any signal in these features is replicated by the bar. This finding validates our methodology for transforming tabular features into images using grayscale bars.

\paragraph{Barttenders are order-invariant} All Barttender metrics reported in Table \ref{tab:chexpert-metrics} are averaged across all six possible order of age, race, and sex. Confidence Intervals (CIs) for these measures, reported in the appendix (Table \ref{tab:chexpert-metrics-all}) are small, indicating that the performance variability between Barttenders with different bar orderings is minimal. This finding provides compelling evidence that the attention on Barttender regions is unaffected by change in order.

\subsection{MIMIC}

\begin{table*}[hbtp]
\floatconts
  {tab:mimic-metrics}
  {\caption{Performance metrics of different approaches trained on various combinations of X-ray images, Image Biomarkers (CTR and CPAR) and 8 Tabular Features to predict Cardiomegaly on a subset of MIMIC ($n=2667$). All reported metrics are averaged across 10-folds.}}
  {\begin{tabular}{llllllllll}
  \toprule
   \bfseries Model        &  \bfseries Method                                 &  \bfseries AUC         &  \bfseries F1          &  \bfseries MCC         \\
  \midrule
   DenseNet-121        & X-ray Image Only                & 0.84 ± 0.05 & 0.84 ± 0.06 & 0.48 ± 0.10 \\
   DenseNet-121        & Blank Barttender                & 0.71 ± 0.01 & 0.79 ± 0.04 & 0.27 ± 0.03 \\
   DenseNet-121        & Image Barttender                & 0.86 ± 0.02 & 0.85 ± 0.03 & 0.51 ± 0.05 \\
   XGBoost            & Image Biomarkers                & 0.85 ± 0.00 & 0.88 ± 0.00 & 0.65 ± 0.01 \\
   XGBoost            & Tabular Data                    & 0.61 ± 0.02 & 0.74 ± 0.02 & 0.21 ± 0.03 \\
   XGBoost            & Image Biomarkers + Tabular Data & 0.83 ± 0.02 & 0.88 ± 0.01 & 0.63 ± 0.03 \\
   Logistic Regression & Image Biomarkers                & 0.88 ± 0.00 & 0.89 ± 0.00 & 0.60 ± 0.01 \\
   Logistic Regression & Tabular Data                    & 0.70 ± 0.00 & 0.82 ± 0.00 & 0.19 ± 0.01 \\
   Logistic Regression & Image Biomarkers + Tabular Data & 0.88 ± 0.00 & 0.89 ± 0.00 & 0.61 ± 0.01 \\
  \bottomrule
  \end{tabular}}
\end{table*}

\paragraph{Barttenders offer a faithful comparison of imaging to non-imaging features} 
Table \ref{tab:mimic-metrics} shows that the relative ranking of model performances in predicting Cardiomegaly from XGBoost and logistic regression are faithfully replicated by the Barttenders — adding image features to tabular features improves the model performance significantly and with similar magnitudes, as expected. 

The negligible difference (0.02) between the Image Barttender and XGBoost with tabular EHR data and image biomarkers shows that our approach can compare imaging and tabular data as accurately as XGBoost/Regression with image biomarkers. Barttender requires only the original imaging and non-imaging data, with no complex pipeline for extracting biomarkers necessary. Thus, our approach facilitates rapid prototyping and evaluation of whether imaging or non-imaging covariates drive prediction.

Interestingly, the Blank Barttender outperforms the XGBoost and logistic regression trained on tabular features on AUC and MCC (within the standard deviation) and on all other metrics (outside the standard deviation). This finding suggests that the deep learning model may be learning non-linear interactions between the bars that shallow models like XGBoost and logistic regression were unable to learn, but this phenomena needs to be studied further.

Table \ref{tab:mimic-metrics} shows a negligible difference between the performance of models with imaging data and models with both imaging and tabular data. This result indicates the tabular features did not have additional predictive information to offer when image information was included. This result inherently makes sense for Cardiomegaly, a condition most commonly diagnosed from images. However, for certain patients, as evidenced from the sample-level IoU, the tabular features played an important role in the model's prediction.


\begin{figure}[htbp]
\floatconts
  {fig:blank}
  {\caption{Raincloud plot of gIoUs for Blank Barttenders computed using saliency maps.}}
  {%
      \includegraphics[width=\linewidth]{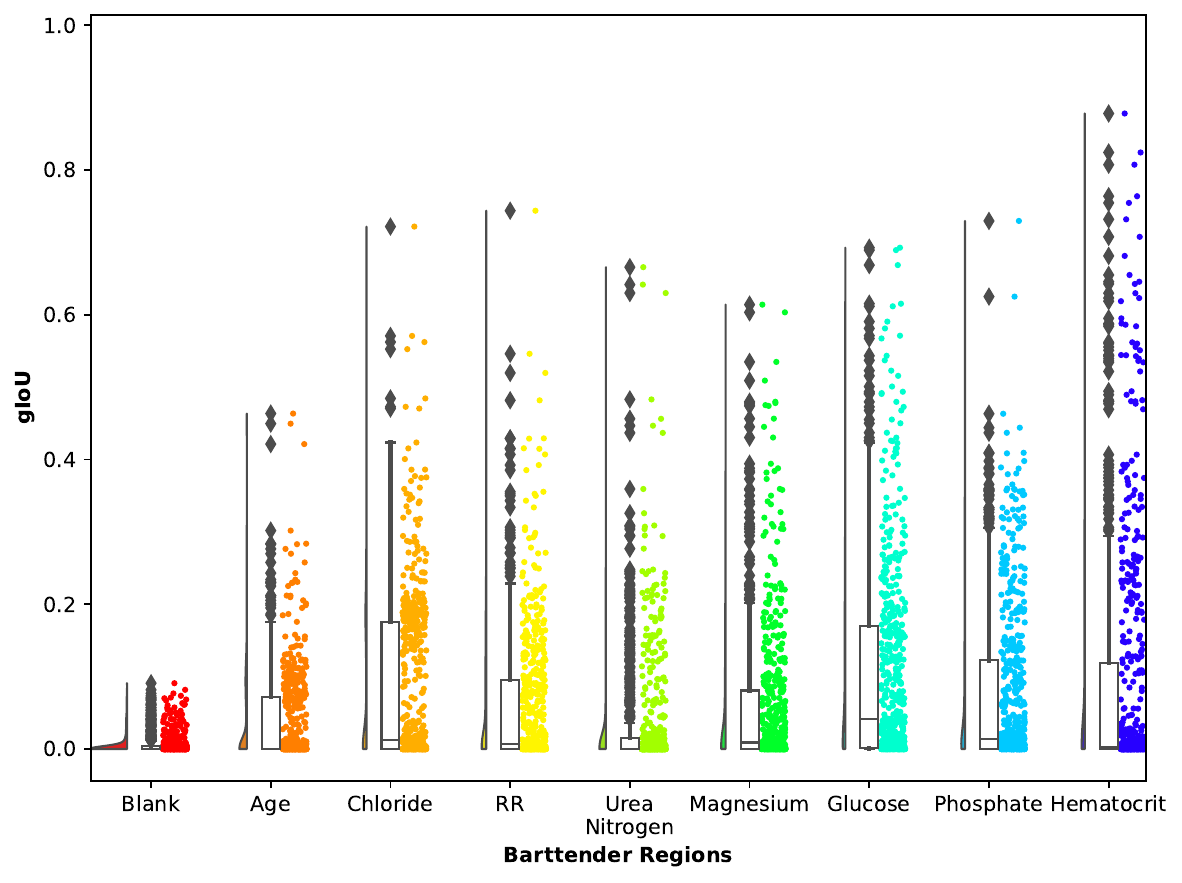}
  }
\end{figure}

Figures \ref{fig:xray} and \ref{fig:blank} offer insights into the distribution of the mean pixel intensity scaled by the IoU across the test set for Image and Blank Barttenders respectively. When the X-ray image is included (Figure \ref{fig:xray}), the median of the distribution across all Barttender bars tends to 0 (the box plot is not visible). However, for all Barttender bars, numerous samples have IoU-scaled mean intensities that are high — some outliers even have IoU-scaled mean intensities comparable to the X-ray. Thus, even though the X-ray has the maximum signal across the test set, the non-imaging features are essential for prediction for certain individuals. This shows the utility of considering sample-level IoUs, as these measures allow us to easily isolate such outliers for further analysis. This high variability for the bars is reflected in the Blank Barttender as well (Figure \ref{fig:blank}), where a significant number of samples have IoU-scaled means higher than the median of the distribution.

Even though the performance of the Barttenders, XGBoost and logistic regression are comparable, their feature importance rankings have significant differences. For the Blank Barttender case in Table \ref{tab:Barttender_comparison_all}, Glucose, Phosphate and Chloride have the highest gIoU values across the entire test set. However, Urea Nitrogen and Age have the lowest gIoU values. This is in contrast to Logistic Regression, where Urea Nitrogen and Age are the top two most important features, while Glucose in the least important, as shown in Table \ref{tab:logistic_regression_feature_importance}. Similarly, Urea Nitrogen is most important for XGBoost when we consider only tabular features, and glucose is least important, as shown in Table \ref{tab:xgboost_feature_importance}. However, both Phosphate and Chloride are among the top 3 most important features for XGBoost, similar to Blank Barttender and in contrast to logistic regression. Similar differences in feature importance appear to be present when both imaging and non-imaging features are considered together. This illustrates that the three models used in our analysis learned different relationships between the tabular features despite performing comparably.

\section{Discussion}
\label{sec:discussion}

In this paper, we introduced \textit{Barttender}, an approachable, customizable, and explainable method for comparing the performance of imaging and non-imaging data on any task using deep learning. Our approach transforms tabular features into grayscale bars, which are appended to corresponding medical images to create the Image Barttender, and also to blank images to create a ``control dataset," called the \textit{Blank Barttender}. A deep learning model of choice is then trained on both the Image and Blank Barttenders, and the relative change in performance is used to evaluate images against non-imaging features. We constructed a robust explainability framework for our approach to assess the relative importance of imaging and non-imaging features by aggregating attribution maps across the test set using gIoU. 

We presented a proof-of-concept for our methodology by evaluating it on two widely used medical imaging datasets — CheXpert and MIMIC. We showed that our approach performs comparably to shallow models trained on scalar representations of the imaging and non-imaging data. This shallow model-based framework, which is most commonly used to compare imaging and non-imaging data, requires domain knowledge and additional training data and steps to extract scalar image biomarkers. Our approach performs comparably out-of-the-box without incorporating any domain knowledge through custom image biomarkers or risk scores. Thus, our approach can significantly speed up the time taken by researchers to complete analyses. Our explainability module was able to deliver key insights into how the deep learning model uses the different features available in the Barttender for prediction. We observed the bars played an outsized role in predicting Cardiomegaly for many individuals in the Image Barttender setting, despite the lower feature importance (gIoU) for tabular features. This provides further evidence for the potential benefit of leveraging multi-modal data at scale in healthcare. We believe such an approach can enhance explainability in clincal settings as well, but to prove this, we would need to execute a study with clinicians, asking the question if, for a given sample, the Image Barttender helps in the interpretation of the prediction.

We believe converting tabular data to images can be advantageous compared to converting medical image data to scalars or embeddings. Conversion of images to scalars limits any analyses to classical models like XGBoost and logistic regression. This leads to an unfair comparison between images and non-imaging data, unless the scalar values are representative of all of the image’s pertinent predictive information. Moreover, embeddings compromise local interpretability. With images, one can review attention maps to understand which parts of the image (e.g., a chest X-ray) the model pays attention to. This may be useful in understanding medical images, especially when explaining the basis for model predictions for a specific patient. In contrast, when converting tabular data into bars, we do not lose any predictive information since we preserve the relative ranking of the scalar through quantile normalization. This ranking maps well to absolute values, since we have a representative and large-scale dataset. We also note that keeping images intact avoids information loss inherent in converting images to less expressive forms like scalars and embeddings. 

Barttenders allow researchers to explore whether images contribute more than traditional risk factors (encoded in non-imaging data) in DL. Typically, separating the influences of tabular covariates from image features in DL is challenging \citep{degrave2021ai}, but our feature importance measures allow researchers to disentangle them. Combined with image attributions, we believe these measures significantly improve the explainability of imaging-based DL solutions for tasks like disease prediction. If adopted in clinical settings, Barttender’s sample-level "IoU" could offer clinicians insights into risk factors influencing model predictions beyond the images. However, we recognize that clinical decision-making is complex, and our framework needs to be studied in clinical contexts to understand its efficacy in practice. 

\subsection{Future work} 

Future work should address limitations of this work. First, a limitation of this work is that evaluating approaches for the prediction of Cardiomegaly has little clinical relevance, as it can be diagnosed by doctors through a simple manual review of the X-ray. Although the image biomarkers developed for Cardiomegaly by \cite{duvieusart_multimodal_2022} for MIMIC made it an ideal setting for our proof-of-concept, future work should target applying our approach in scenarios that have greater clinical relevance. Second, our gIoU measure does not have the same statistical and causal foundations as feature importances for models such as regression. In linear regression modeling, researchers make causal interpretations by interpreting the beta coefficients or weights for each input variable; e.g., a change in the predictor and its impact on the outcome is interpreted with the weight or beta coefficient. However, researchers cannot make casual claims with gIoU — they can only posit correlations. Furthermore, we also note that higher dimensional data like time-series data or similarly complex data may not be suited to such a representation and use case for interpretability. What we advise is that data-driven “feature selection” be used as a first step prior to including variables as bars, as explained in the “Feature Selection” subsection in Section 4.2.

A promising direction for future work would be to apply this methodology to diseases where imaging data is not yet utilized in practice to better understand the potential value of incorporating imaging data. For example, it could be applied to type-II diabetes, where tabular risk factors like glucose are sufficient to diagnose disease, but other types of information such as complex imaging (MRI) could be beneficial to better understand risk for complications. Comparing existing risk score models like the Framingham Risk Score with cardiovascular imaging data could be another interesting application area for cardiovascular diseases.

\newpage

\bibliography{main}

\appendix

\section{Related Works}
\label{apd:related_works}
There are numerous approaches to combining non-imaging and image modalities for disease prediction. \citep{huang_fusion_2022}. Investigators have trained multimodal deep learning fusions of medical imaging and non-imaging data (e.g., tabular) for disease prediction \citep{holste_end--end_2021,hager_best_2023,cui_deep_2023,huang_fusion_2022}. Some of these approaches can be used to compare the performance between medical imaging and non-imaging data. For example, one approach relies on a self-supervised contrastive learning framework that leverages imaging and tabular data for predicting cardiovascular risks \citep{hager_best_2023}. For explainability, \cite{hager_best_2023} utilized Integrated Gradients \citep{sundarajan_axiomatic_2017} to quantify the importance of individual tabular features in generating mutli-model embeddings. A limitation includes the complexity of obtaining joint embeddings and the inability to encode data missingness. For missing data, their approach relies on multivariate imputation. This imputation scheme relies on assuming that data are missing at random (MAR), but clinical tabular data often contain missing not at random (MNAR) features, leading to potentially biased results \citep{van_walraven_imputing_2023,li_imputation_2021}. 

Other investigators have encoded tabular data in images through deep learning approaches, achieving model performance at par or beyond simpler machine learning approaches like XGBoost \citep{zhu_converting_2021,medeiros_neto_comparative_2023,
damri_towards_2024,alkhodari_circadian_2024,huang_fusion_2022}. However, these encoding methods are often complex, and thus not explainable on the image-level. 

Simpler methods to encode non-imaging data into images can allow researchers to obtain sample-level explanations and work towards global-level explanations. For example, \cite{alkhodari_circadian_2024} created  a ``dartboard" image from tabular and time-series features to predict heart variability. Each ring of the dartboard contains a static tabular or scalar feature--for instance, age, sex, ethnicity--or a real-valued continuous variable, like body mass index (BMI). Colors and their intensity were applied based on the  feature values. To encode heart-rate variability (HRV), a time-varying feature across a 24-hour cycle, the dartboard was split into 24 equal sections. These sections were re-scaled based on the HRV values for their respective hour.  Using a deep learning model trained on these images, average saliency maps across the entire test set for each label were quantified to assess feature importance.

We propose a new method to compare imaging and non-imaging data by embedding tabular data as bars,
\textit{Barttender}, which are appended to each medical image. Both the dartboard representation and the bars in Barttenders are similar – both convert tabular covariates into rings or bars, respectively. We expand the idea into a generalized framework, utilizing bars of varying color intensity to compare imaging and non-imaging data within the deep learning space. We believe this use is novel and goes beyond the original approach — \cite{alkhodari_circadian_2024}, which did not explore multimodal representations. We also develop generalized rules for converting both categorical and continuous variables into bars, as well as a global measure of feature importances, gIoU.

\section{Experiments}

\subsection{Preparing the multi-modal MIMIC Cardiomegaly sample}
\label{sec:mimic_prep}
This multi-modal sample combines data from the MIMIC-IV and MIMIC-CXR-JPG datasets \citep{johnson_mimic-iv_nodate, johnson_mimic-cxr-jpg_nodate}, preprocessed as directed by \cite{duvieusart_multimodal_2022} (details can be found in the appendix, section \ref{sec:mimic_prep}).  
MIMIC-IV is a large, publically available, and de-identified EHR dataset of over $265,000$ patients admitted to the Beth Israel Deaconess Medical Center in Boston, MA \citep{johnson_mimic-iv_nodate}. The MIMIC-CXR-JPG dataset contains chest X-ray scans for a large subset of the MIMIC-IV patients \citep{johnson_mimic-cxr-jpg_nodate}. To prepare the multi-modal Cardiomegaly dataset from \cite{duvieusart_multimodal_2022}, we first removed all \textit{uncertain} and \textit{no mention} Cardiomegaly labels and kept only posterior-anterior (PA) X-rays to avoid the unnatural enlargement of the cardiac silhouette that can occur in other views. We then linked ICU stays from MIMIC-IV to the closest X-ray study from those remaining, within a window of 365 days before ICU admission and up to 90 days post-discharge. This produced a dataset consisting of $2667$ samples across $X$ patients, where each sample contained a chest X-ray, values for $14$ laboratory tests, values for $14$ ICU chart/vital sign values, and patient metadata, including demographics. For further details, please refer to \cite{duvieusart_multimodal_2022}.

\subsection{General Experimental Setup}

We used a DenseNet-121 model pre-trained on ImageNet for all Barttender and image-based experiments. All images in our experiments are 224x224, and the bars comprise of about 20\% of the image across all experiments. We used DenseNet-121 since it is used extensively in medical image analysis, and widely regarded as a powerful yet tractable CNN architecture \citep{irvin_chexpert_2019}. We trained all models using 10-fold cross-validation with early stopping, and employed train-validation-test splits of 70\%, 10\%, and 20\%, respectively. The test samples were held out across all folds.

To compute image attributions for the DenseNet model used in our experiments, we used two approaches — classic gradient-based saliency maps and Integrated Gradients. In the classic gradient-based saliency map approach, we computed the gradient of the model’s output with respect to the input image pixels. This gradient highlights which pixels have the most influence on the model’s prediction, providing a straightforward but effective way to visualize regions of the image that contribute to the model's decision. In Integrated Gradients, we approximated the integration of the gradients of the encoder along the straightline path from a baseline sample to the test sample in question. This integration yields an intensity for the pixel attribution, quantifying the pixel's relevance to the downstream prediction for that sample. As discussed in Section \ref{sec:methods}, using multiple approaches for computing image attributions ensured that our findings are generalizable and not dependent on a single attribution method.

\section{Results}

\begin{table*}[htbp]
\setlength{\tabcolsep}{3pt}
\floatconts
    {tab:Barttender_comparison_all}
    {\caption{Comparison of Barttender Feature Importance (computed using $gIoU$) across different regions of X-ray and Blank Barttender for Integrated Gradients. ``Cases" refer to samples that are positive for Cardiomegaly, and "Controls" refer to samples that are negative for Cardiomegaly. Computing the gIoUs for cases and controls separately allows us to infer the relative importance of a feature }}
    {\begin{tabular}{lcccc|cccc}
    \toprule
    \multirow{2}{*}{ \bfseries Barttender Region} & \multicolumn{4}{c}{ \bfseries Blank Bartt., Integrated Gradients} & \multicolumn{4}{c}{ \bfseries X-Ray Bartt., Integrated Gradients} \\
    \cline{2-5} \cline{6-9}
     & All & Cases & Controls & Cases - Controls & All & Cases & Controls & Cases - Controls \\
    \midrule
Image (X-Ray/Blank)  & 0.0040  & 0.0040  & 0.0030  & 6.0e-04  & 0.0700  & 0.0710  & 0.0700  & 0.0010 \\
Chloride Bar         & 0.0660  & 0.0770  & 0.0410  & 0.0360   & 0.0013  & 0.0013  & 7.0e-04 & 6.0e-04 \\
Phosphate Bar        & 0.1100  & 0.0970  & 0.1300  & -0.0340  & 0.0011  & 0.0012  & 0.0010  & 2.0e-04 \\
Hematocrit Bar       & 0.1100  & 0.1000  & 0.1300  & -0.0270  & 9.2e-04 & 8.7e-04 & 0.0010  & -1.3e-04 \\
Resting Rate Bar     & 0.0510  & 0.0580  & 0.0350  & 0.0230   & 5.7e-04 & 5.2e-04 & 6.7e-04 & -1.5e-04 \\
Age Bar              & 0.0270  & 0.0320  & 0.0150  & 0.0170   & 0.0015  & 0.0017  & 0.0010  & 7.0e-04 \\
Glucose Bar          & 0.1200  & 0.1270  & 0.1200  & -0.0070  & 0.0049  & 0.0047  & 0.0053  & -6.0e-04 \\
Magnesium Bar        & 0.0580  & 0.0570  & 0.0600  & -0.0040  & 0.0013  & 0.0015  & 0.0010  & 5.0e-04 \\
Urea Nitrogen Bar    & 0.0320  & 0.0310  & 0.0340  & -0.0030  & 3.0e-04 & 3.0e-04 & 1.0e-04 & 2.0e-04 \\
    \bottomrule
    \end{tabular}}
\end{table*}

\begin{table}[htbp]
\setlength{\tabcolsep}{3pt}
\floatconts
    {tab:logistic_regression_feature_importance}
    {\caption{Logistic Regression Feature Importance across Different Runs. This table combines Z-scores for feature importance from tabular data, image biomarkers, and their combination.}}
    {\begin{tabular}{lccc}
    \toprule
    \bfseries Feature & \bfseries Tabular & \bfseries \Centerstack{Tabular \\ + \\ Image} & \bfseries Image \\
    \midrule
    CTR            &  -   & 6.53 & 8.09 \\
    CPAR           &  -   & 6.66 & 6.36 \\
    Age            & 6.83 & 3.24 &  - \\
    Urea Nitrogen  & 3.36 & 1.53 &  - \\
    Magnesium      & 2.78 & 0.02 &  - \\
    Chloride       & 2.44 & 1.08 &  - \\
    Phosphate      & 2.05 & 0.86 &  - \\
    Hematocrit     & 1.56 & 0.96 &  - \\
    Resting Rate   & 0.47 & 1.06 &  - \\
    Glucose        & 0.33 & 0.01 &  - \\
    \bottomrule
    \end{tabular}}
\end{table}

\begin{table}[htbp]
\setlength{\tabcolsep}{3pt}
\floatconts
    {tab:xgboost_feature_importance}
    {\caption{XGBoost Feature Importance for Various Runs. The table combines feature importance values from three different runs: Tabular Data, Image Biomarkers, and a combination of both.}}
    {\begin{tabular}{lccc}
    \toprule
    \bfseries Feature & \bfseries Tabular & \bfseries \Centerstack{Tabular \\ + \\ Image} & \bfseries Image \\
    \midrule
    CTR            & -    & 0.57 & 0.23 \\
    CPAR           & -    & 0.42 & 0.17 \\
    Urea Nitrogen  & 0.18 & -    & 0.08 \\
    Phosphate      & 0.12 & -    & 0.06 \\
    Chloride       & 0.12 & -    & 0.06 \\
    Age            & 0.11 & -    & 0.06 \\
    Magnesium      & 0.11 & -    & 0.07 \\
    Resting Rate   & 0.11 & -    & 0.07 \\
    Hematocrit     & 0.11 & -    & 0.07 \\
    Glucose        & 0.10 & -    & 0.07 \\
    \bottomrule
    \end{tabular}}
\end{table}

\begin{table*}[htbp]
\floatconts
  {tab:chexpert-metrics-all}
  {\caption{Average performance metrics for DenseNet-121 across different feature orderings. Each value represents the average of results from 10-fold cross-validation runs.}}
{\begin{tabular}{llccc}
\toprule
\bfseries Model          & \bfseries Label                        & \bfseries AUC     & \bfseries F1    & \bfseries MCC  \\
\midrule
    Blank Barttender & No Finding                   & 0.50 ± 0.00 & 0.46 ± 0.00 & 0.00 ± 0.00 \\
    Blank Barttender & Atelectasis                   & 0.51 ± 0.00 & 0.41 ± 0.01 & 0.03 ± 0.01 \\
    Blank Barttender & Cardiomegaly                   & 0.53 ± 0.01 & 0.48 ± 0.01 & 0.10 ± 0.01 \\
    Blank Barttender & Consolidation                 & 0.53 ± 0.00 & 0.50 ± 0.01 & 0.06 ± 0.00 \\
    Blank Barttender & Edema                        & 0.53 ± 0.00 & 0.48 ± 0.01 & 0.07 ± 0.01 \\
    Blank Barttender & Enlarged Cardiomediastinum     & 0.50 ± 0.01 & 0.42 ± 0.01 & 0.01 ± 0.01 \\
    Blank Barttender & Fracture                     & 0.50 ± 0.00 & 0.42 ± 0.00 & 0.00 ± 0.00 \\
    Blank Barttender & Lung Lesion                  & 0.50 ± 0.00 & 0.46 ± 0.00 & 0.00 ± 0.00 \\
    Blank Barttender & Lung Opacity                 & 0.50 ± 0.00 & 0.42 ± 0.00 & 0.00 ± 0.00 \\
    Blank Barttender & Pleural Effusion             & 0.53 ± 0.01 & 0.49 ± 0.03 & 0.06 ± 0.01 \\
    Blank Barttender & Pleural Other                & 0.50 ± 0.00 & 0.47 ± 0.00 & 0.00 ± 0.00 \\
    Blank Barttender & Pneumonia                    & 0.50 ± 0.00 & 0.43 ± 0.00 & 0.00 ± 0.00 \\
    Blank Barttender & Pneumothorax                 & 0.50 ± 0.00 & 0.47 ± 0.00 & 0.00 ± 0.00 \\
    Blank Barttender & Support Devices              & 0.51 ± 0.00 & 0.42 ± 0.01 & 0.05 ± 0.02 \\
    Logistic Regression & No Finding                   & 0.58 ± 0.00 & 0.23 ± 0.00 & 0.08 ± 0.00 \\
    Logistic Regression & Atelectasis                   & 0.55 ± 0.00 & 0.35 ± 0.01 & 0.05 ± 0.00 \\
    Logistic Regression & Cardiomegaly                   & 0.57 ± 0.00 & 0.38 ± 0.01 & 0.07 ± 0.00 \\
    Logistic Regression & Consolidation                & 0.54 ± 0.00 & 0.31 ± 0.01 & 0.03 ± 0.00 \\
    Logistic Regression & Edema                         & 0.49 ± 0.00 & 0.22 ± 0.01 & -0.05 ± 0.00 \\
    Logistic Regression & Enlarged Cardiomediastinum & 0.59 ± 0.00 & 0.43 ± 0.01 & 0.10 ± 0.00 \\
    Logistic Regression & Fracture                     & 0.47 ± 0.00 & 0.19 ± 0.00 & -0.03 ± 0.00 \\
    Logistic Regression & Lung Lesion                   & 0.53 ± 0.00 & 0.18 ± 0.01 & 0.02 ± 0.00 \\
    Logistic Regression & Lung Opacity                  & 0.57 ± 0.00 & 0.40 ± 0.01 & 0.06 ± 0.00 \\
    Logistic Regression & Pleural Effusion             & 0.53 ± 0.00 & 0.34 ± 0.01 & 0.05 ± 0.00 \\
    Logistic Regression & Pleural Other                & 0.47 ± 0.00 & 0.11 ± 0.00 & -0.02 ± 0.00 \\
    Logistic Regression & Pneumonia                    & 0.50 ± 0.00 & 0.20 ± 0.00 & -0.01 ± 0.00 \\
    Logistic Regression & Pneumothorax                 & 0.58 ± 0.00 & 0.19 ± 0.00 & 0.05 ± 0.00 \\
    Logistic Regression & Support Devices              & 0.46 ± 0.00 & 0.20 ± 0.01 & -0.10 ± 0.00 \\
    Image Barttender    &   No Finding                 & 0.78 ± 0.02 & 0.78 ± 0.01 & 0.58 ± 0.02 \\
    Image Barttender    &   Atelectasis                & 0.82 ± 0.01 & 0.82 ± 0.01 & 0.64 ± 0.02 \\
    Image Barttender    &   Cardiomegaly               & 0.80 ± 0.01 & 0.80 ± 0.01 & 0.61 ± 0.01 \\
    Image Barttender    &   Consolidation              & 0.81 ± 0.01 & 0.81 ± 0.01 & 0.64 ± 0.01 \\
    Image Barttender    &   Edema                      & 0.78 ± 0.01 & 0.78 ± 0.01 & 0.57 ± 0.01 \\
    Image Barttender    &   Enlarged Cardiomediastinum  & 0.81 ± 0.01 & 0.81 ± 0.01 & 0.64 ± 0.01 \\
    Image Barttender    &   Fracture                   & 0.68 ± 0.01 & 0.69 ± 0.01 & 0.43 ± 0.01 \\
    Image Barttender    &   Lung Lesion                & 0.62 ± 0.01 & 0.65 ± 0.01 & 0.37 ± 0.02 \\
    Image Barttender    &   Lung Opacity               & 0.84 ± 0.01 & 0.84 ± 0.01 & 0.68 ± 0.01 \\
    Image Barttender    &   Pleural Effusion           & 0.82 ± 0.01 & 0.81 ± 0.01 & 0.64 ± 0.01 \\
    Image Barttender    &   Pleural Other              & 0.61 ± 0.01 & 0.64 ± 0.01 & 0.34 ± 0.02 \\
    Image Barttender    &   Pneumonia                  & 0.67 ± 0.02 & 0.68 ± 0.02 & 0.42 ± 0.02 \\
    Image Barttender    &   Pneumothorax               & 0.61 ± 0.03 & 0.63 ± 0.03 & 0.34 ± 0.04 \\
    Image Barttender    &   Support Devices            & 0.76 ± 0.01 & 0.76 ± 0.01 & 0.53 ± 0.01 \\
\bottomrule
\end{tabular}}
\end{table*}


\end{document}